\documentclass[12pt]{article}
\pdfoutput=1

\usepackage[bulletsep]{collref}
\usepackage{amssymb,graphicx}
\usepackage[intlimits]{amsmath}
\usepackage{bbm}
\usepackage[small]{subfigure}

\usepackage{MnSymbol}

\makeatletter \@addtoreset{equation}{section} \makeatother

\makeatletter
\let\old@startsection=\@startsection
\let\oldl@section=\l@section
\renewcommand{\@startsection}[6]{\old@startsection{#1}{#2}{#3}{#4}{#5}{#6\mathversion{bold}}}
\renewcommand{\l@section}[2]{\oldl@section{\mathversion{bold}#1}{#2}}
\makeatother

\makeatletter
\let\old@makecaption=\@makecaption
\def\@makecaption{\small\old@makecaption}
\makeatother

\renewcommand{\geq}{\geqslant}

\begin{document}


\thispagestyle{empty}
\begin{flushright}\footnotesize
\texttt{NORDITA 2020-022}
\vspace{0.6cm}
\end{flushright}

\renewcommand{\thefootnote}{\fnsymbol{footnote}}
\setcounter{footnote}{0}

\begin{center}
{\Large\textbf{\mathversion{bold} Quiver CFT at strong coupling }
\par}

\vspace{0.8cm}

\textrm{K.~Zarembo\footnote{Also at ITEP, Moscow, Russia}}
\vspace{4mm}

\textit{Nordita, KTH Royal Institute of Technology and Stockholm University,
Roslagstullsbacken 23, SE-106 91 Stockholm, Sweden}\\
\textit{Niels Bohr Institute, Copenhagen University, Blegdamsvej 17, 2100 Copenhagen, Denmark}\\
\vspace{0.2cm}
\texttt{zarembo@nordita.org}

\vspace{3mm}


\par\vspace{1cm}

\textbf{Abstract} \vspace{3mm}

\begin{minipage}{13cm}
The circular Wilson loop in the two-node quiver CFT is computed at large-$N$ and strong 't~Hooft coupling by solving the localization matrix model.
\end{minipage}
\end{center}

\vspace{0.5cm}


\setcounter{page}{1}
\renewcommand{\thefootnote}{\arabic{footnote}}
\setcounter{footnote}{0}

\section{Introduction}

An $SU(N_c)$ gauge theory with $N_f=2N_c$ fundamental hypermultiplets, often called super-QCD, is perhaps the simplest $\mathcal{N}=2$ superconformal theory. Since its conformal anomaly does not satisfy $a=c$, a putative  holographic dual must always remain stringy, no matter how large the 't~Hooft coupling is \cite{Gadde:2009dj}, in contradistinction, for instance, to $\mathcal{N}=4$ super-Yang-Mills (SYM). In spite of this striking difference, SQCD and SYM are connected by a family of superconformal theories, all having weakly-coupled duals. It would be interesting to understand how the string description breaks down or becomes strongly-coupled at the SQCD point.

\begin{figure}[t]
\begin{center}
 \centerline{\includegraphics[width=5cm]{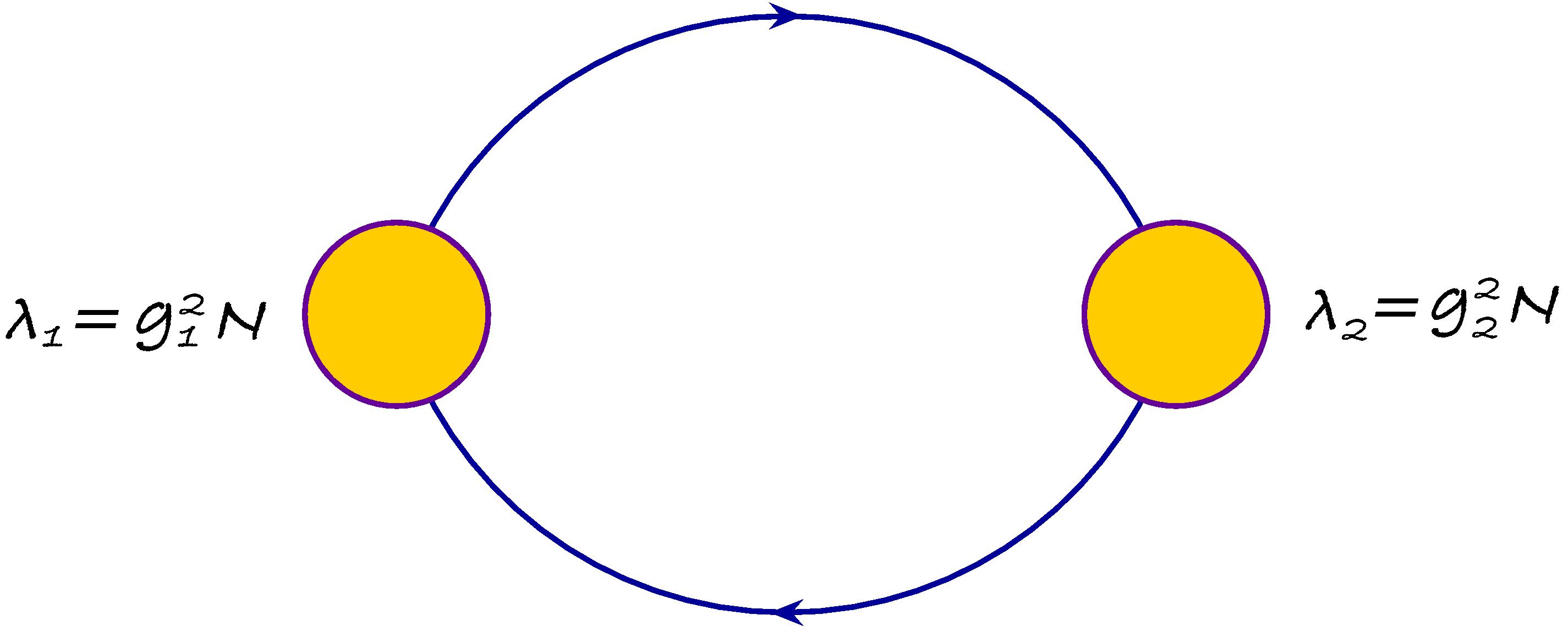}}
\caption{\label{qui}\small Two-node quiver.}
\end{center}
\end{figure}

The interpolating theory is obtained by gauging the flavor group of SQCD. The result is an  $SU(N)\times SU(N)$  quiver with bi-fundamental matter and two independent couplings (fig.~\ref{qui}). Once flavor gauge fields decouple at $\lambda _2=0$, the quiver becomes equivalent to SQCD augmented with a free vector multiplet that restores $a=c$. For equal couplings, the symmetry is enhanced by an extra $\mathbbm{Z}_2$. This is not accidental, as at $\lambda _1=\lambda _2$ the quiver is equivalent to the $\mathbbm{Z}_2$ orbifold of $\mathcal{N}=4$ SYM \cite{Lawrence:1998ja}. The orbifold and the parent  SYM theory share the same planar diagrams \cite{Bershadsky:1998cb} and hence are equivalent at $N\rightarrow \infty $. 

The holographic dual of the quiver is string theory on the $AdS_5\times (S^5/\mathbbm{Z}_2)$ orbifold  \cite{Kachru:1998ys}, where $\mathbbm{Z}_2$ acts by flipping the four coordinates of $S^5$ in the $\mathbbm{R}^6$ embedding, reflecting the 2+4 split of the $\mathcal{N}=4$ scalars between the vector and  hypermultiplet of $\mathcal{N}=2$. 

The vastly different strong-coupling behavior of SYM and SQCD manifests itself in the expectation value of the circular Wilson loop, which can be computed from first principles  in both cases using localization \cite{Pestun:2007rz}.  The  SYM Wilson loop nicely exponentiates \cite{Erickson:2000af,Drukker:2000rr}:
\begin{equation}\label{N=4WL}
 W_{\rm SYM}=\frac{2}{\sqrt{\lambda }}\,I_1\left(\sqrt{\lambda }\right)
 \stackrel{\lambda \rightarrow \infty }{\simeq} \sqrt{\frac{2}{\pi }}\,\lambda ^{-\frac{3}{4}}\,{\rm e}\,^{\sqrt{\lambda }},
\end{equation}
in agreement with the minimal area law in $AdS_5$. Indeed, the regularized area of the circle is $-2\pi $ \cite{Berenstein:1998ij,Drukker:1999zq}, the string tension is 
\begin{equation}\label{String-Ten}
 T=\frac{\sqrt{\lambda }}{2\pi }\,.
\end{equation}
Together they give $\sqrt{\lambda }$ in the exponent. 

The Wilson loop in the quiver CFT also exponentiates, in terms of the effective coupling \cite{Rey:2010ry}:
\begin{equation}\label{eff-coupling}
 \frac{2}{\lambda }=\frac{1}{\lambda _1}+\frac{1}{\lambda _2}\,,
\end{equation}
in accord with expectations from AdS/CFT, as exactly the same coupling
controls the string tension \cite{Lawrence:1998ja,Klebanov:1999rd,Gadde:2010zi}, while the minimal surface is unaffected by the orbifold projection. The notion of effective coupling actually applies to a larger class of $\mathcal{N}=2$ superconformal theories and goes beyond the strong-coupling regime \cite{Mitev:2014yba,Mitev:2015oty}.

On the contrary,  in SQCD the Wilson loop does not exponentiate (we denote the SQCD 't~Hooft coupling by $\lambda _1$, keeping in mind its embedding in the quiver) \cite{Passerini:2011fe}:
\begin{equation}\label{WSQCD}
 W_{\rm SQCD}\stackrel{\lambda _1\rightarrow \infty }{\simeq} \,{\rm const}\,\frac{\lambda_1 ^3}{\left(\ln\lambda_1 \right)^{\frac{3}{2}}}\,.
\end{equation}
Such a power+log behavior is hardly consistent with a semi-classical string interpretation.

To the leading order the Wilson loop only depends on the average of the inverse couplings. The difference does not show up in the exponent. In string theory, the difference defines a theta-angle on the 
worldsheet  \cite{Lawrence:1998ja,Klebanov:1999rd,Gadde:2010zi}:
\begin{equation}\label{theta}
 \theta =\pi -\pi \,\frac{\frac{1}{\lambda _1}-\frac{1}{\lambda _2}}{\frac{1}{\lambda _1}+\frac{1}{\lambda _2}}=\frac{2\pi \lambda _1}{\lambda _1+\lambda _2}\,.
\end{equation}
Proper definition of the corresponding term in the string action requires resolution of the orbifold singularity. Supersymmetry-preserving resolution involves a non-contractable two-cycle collapsing to zero size when regularization is removed. The theta-term  measures the wrapping number of the worldsheet around this non-contractable cycle.  Interestingly, the symmetric point ($\lambda _1=\lambda _2$) corresponds to the $\pi$-flux  ($\theta =\pi $) and not zero as one could possibly expect. The theta-term breaks CP such that interchanging the two gauge groups ($\lambda _1\leftrightarrow\lambda _2$) entails a parity transformation on the worldsheet: $\theta \rightarrow 2\pi -\theta $.

This wonderful picture calls for a quantitative test. A first-principles string calculation would be particularly interesting.
This is not what we will do here. Instead we will explore the circular Wilson loop in the stringy regime, but by purely field-theoretic methods, namely by solving the localization matrix model \cite{Pestun:2007rz} to the first order in the strong-coupling expansion, expending the results  in \cite{Rey:2010ry} beyond the leading exponential. The leading order does not carry any theta-dependence and the Wilson loop expectation value is essentially the same as in SYM. The "one-loop" correction we are going to compute can serve as a testbed for string theory on the orbifold with the B-flux along with the spectral data known in quite a detail at any coupling \cite{Gadde:2009dj,Gadde:2010zi,Gadde:2010ku}.

\section{Localization}

The field content of the $SU(N)\times SU(N)$ quiver consists of two vectors multiplets in the adjoint\footnote{Only bosonic fields are displayed.}: $(A_{a\mu} ,\Phi _a,\Phi '_a)$,  $a=1,2$, and bi-fundamental  matter: $(X,Y,X^\dagger ,Y^\dagger )$: $D_\mu X=\partial _\mu X+A_{1\mu} X-XA_{2\mu} $. We will be interested in the Wilson loop expectation value
\begin{equation}
 W_a=\left\langle \frac{1}{N}\,{\rm P}\exp
 \left[\oint_C ds\,\left(i\dot{x}^\mu A_{a\mu} +|\dot{x}|\Phi _a\right) \right]
 \right\rangle,
\end{equation}
for the circular contour $C$. 

After the theory is placed on the four-sphere the problem reduces to a finite-dimensional matrix integral over zero modes of the vector-mul\-tip\-let sca\-lars. In the eigenvalue representation, $\Phi _a=\mathop{\mathrm{diag}}(a_{a1}\ldots a_{aN})$, the localization integral is \cite{Pestun:2007rz}:
\begin{equation}\label{localizationMM}
 Z=\int_{}^{}\prod_{a=1}^{2}\prod_{i}^{}da_{ai}\,\,
 \frac{\prod\limits_{a}^{}\prod\limits_{i<j}^{}(a_{ai}-a_{aj})^2H^2(a_{ai}-a_{aj})}{\prod_{ij}^{}H^2(a_{1i}-a_{2j})}\,
 \,{\rm e}\,^{-\sum_{a}^{}\frac{8\pi ^2N}{\lambda _a}\sum_{i}^{}a_{ai}^2},
\end{equation}
where $H(x)$ admits a product representation:
\begin{equation}
 H(x)=\prod_{n=1}^{\infty }\left(1+\frac{x^2}{n^2}\right)^n\,{\rm e}\,^{-\frac{x^2}{n}}.
\end{equation}

The circular Wilson loops correspond to simple exponentials in the localization matrix model:
\begin{equation}
 W_a=\left\langle \frac{1}{N}\sum_{i}^{}\,{\rm e}\,^{2\pi a_{ai}}\right\rangle.
\end{equation}
In contradistinction to $\mathcal{N}=4$ SYM, where the matrix model  is Gaussian \cite{Erickson:2000af,Drukker:2000rr}, the quiver matrix integral is interacting even at the orbifold point $\lambda _1=\lambda _2$. This demonstrates very clearly that the orbifold equivalence is a dynamical phenomenon and only holds in the strict large-$N$ limit. Even at large-$N$ equivalence to the Gaussian model  is not immediately obvious. It can be formally established by inspecting the large-$N$ saddle-point equations.

When written in terms of the the eigenvalue densities, 
\begin{equation}
 \rho _a(x)=\left\langle \frac{1}{N}\,\sum_{i}^{}\delta (x-a_{ai})\right\rangle,
\end{equation}
 the saddle-point equations \cite{Brezin:1977sv} become
\begin{eqnarray}
 \strokedint_{-\mu _1}^{\mu _1}dy\,\rho _1(y)\left(\frac{1}{x-y}-K(x-y)\right)
 +\int_{-\mu _2}^{\mu _2}dy\,\rho _2(y)K(x-y)&=&\frac{8\pi ^2}{\lambda _1}\,x
 \label{inteq1}
\\
\strokedint_{-\mu _2}^{\mu _2}dy\,\rho _2(y)\left(\frac{1}{x-y}-K(x-y)\right)
 +\int_{-\mu _1}^{\mu _1}dy\,\rho _1(y)K(x-y)&=&\frac{8\pi ^2}{\lambda _2}\,x,
\label{inteq2}
\end{eqnarray}
where
\begin{equation}
 K(x)=-\frac{H'(x)}{H(x)}=x\left(\psi (1+ix)+\psi (1-ix)+2\gamma \right).\end{equation}
 The Wilson loops are given by 
\begin{equation}\label{wils-int}
 W_a=\int_{-\mu _a}^{\mu _a}dx\,\rho_a (x)\,{\rm e}\,^{2\pi x}.
\end{equation}
This setup has been used to study Wilson loops in SQCD and quiver CFT, mostly at weak coupling \cite{Fiol:2015mrp,Mitev:2015oty,Billo:2018oog,Billo:2019fbi}. The leading-order strong-coupling solution of the saddle-point equations was obtained in \cite{Rey:2010ry}. We will extend it to the next order in $1/\sqrt{\lambda }$.

When $\lambda _1=\lambda _2=\lambda $, the equations are consistent with the symmetric ansatz $\rho _1=\rho _2$, for which the $K$-terms cancels and one is left with the saddle-point equation of the Gaussian matrix model whose solution is the Wigner semicircle:
\begin{equation}
 \rho (x)=\frac{2}{\pi \mu ^2}\,\sqrt{\mu ^2-x^2}
\end{equation}
with
\begin{equation}
 \mu =\frac{\sqrt{\lambda }}{2\pi }\,.
\end{equation}
This is how orbifold equivalence operates at large $N$.

As observed in  \cite{Rey:2010ry} the semicircular distribution is a good approximation even for unequal $\lambda _1$, $\lambda _2$, provided that both couplings are large and comparable in magnitude. The argument goes as follows. The saddle-point equations reflect the balance of forces between eigenvalues. The $1/(x-y)$ repulsion smoothens the distribution on short scales but dies out at large distances. The external linear force confines the eigenvalues to a finite interval but at  strong coupling is only operative at very large $x$. The bulk of the distribution is thus controlled by the two-body forces mediated by $K(x-y)$. The function $K(x)$ is overall positive and grows as $x\ln x$ at large $x$. As a result, the like eigenvalues attract, while the opposite eigenvalues repel with a force that grows with distance. To balance this force and prevent large terms appearing in the integral equations, the two eigenvalue distributions "lock" making the densities $\rho _{1,2}$ approximately equal. The locking cancels large terms with $K(x-y)$. The cancellation is only approximate in each of the equations  (\ref{inteq1}) and (\ref{inteq2}), but an almost perfect cancellation occurs in their sum \cite{Rey:2010ry}. Thus $\rho _{1}\approx \rho _{2}$ implies that both densities are given by the Wigner distribution whose width is determined by the effective coupling  (\ref{eff-coupling}).

\begin{figure}[t]
\begin{center}
 \centerline{\includegraphics[width=8cm]{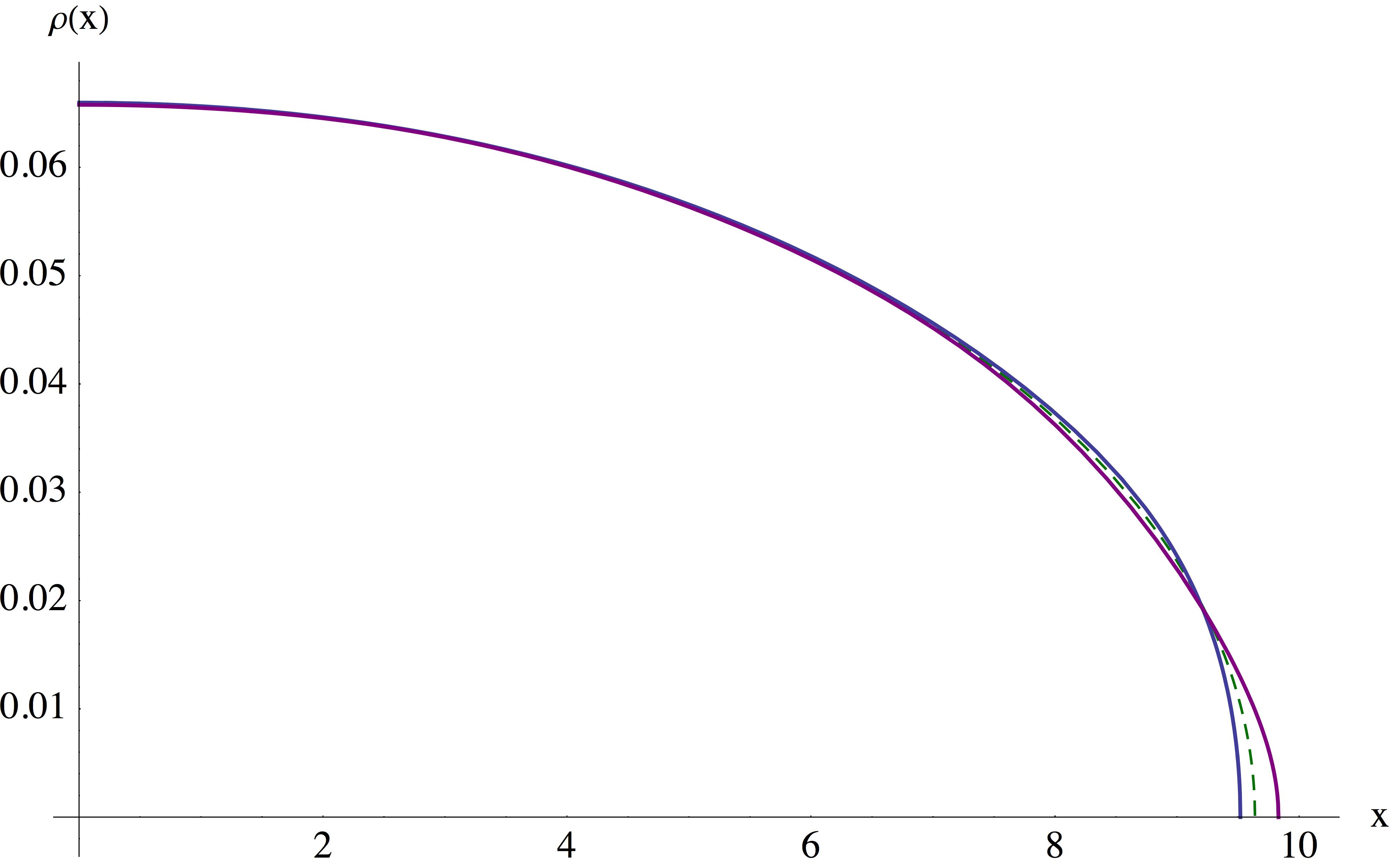}}
\caption{\label{densities-fig}\small The eigenvalue densities $\rho _1$ (purple line) and $\rho _2$ (blue line) obtained by numerically solving (\ref{inteq1}), (\ref{inteq2}) for $\lambda _1=5320$, $\lambda _2=2797$. The dashed line is the Wigner distribution with the effective coupling $\lambda =3667$. The density for the  gauge group with a larger coupling ($\rho _1$) tends to spread more because the restoring force is weaker, hence $\mu _1>\mu_2$, but in  spite of considerable disparity in the coupling strength the difference between $\rho  _1$ and $\rho  _2$ is very small. This is the locking effect. The difference is most pronounced near the spectral edge.}
\end{center}
\end{figure}

This picture agrees very well with numerics (fig.~\ref{densities-fig}). The two densities are approximately the same and deviate from the Wigner distribution only near the spectral edge. But Wilson loops are controlled precisely by the edge, because of their exponential dependence on the eigenvalues. We thus need to know the edge behavior of the densities in detail.
 
 Since $\mu _{1,2}$ are large the Wilson loop exponentiates at strong coupling, as in the SYM, but with a different prefactor determined by the structure of the eigenvalue density near the endpoint. Exactly the same behavior was found in the $\mathcal{N}=2^*$ theory \cite{Chen:2014vka}, where the leading order solution is approximately Gaussian \cite{Buchel:2013id}, while the first strong-coupling correction is determined by a fairly complicated boundary dynamics. We conjecture that these features are common to
all $\mathcal{N}=2$ theories with weakly-coupled holographic duals.
 Wigner density in the bulk is accompanied by $\mathcal{O}(1)$ deviations at the edge. As in  \cite{Chen:2014vka} we will solve the integral equations in two steps, first in the bulk and then at the boundary, matching the two solutions in their overlapping regime of validity.

\section{Bulk}

It does not make sense to plug $\rho _1=\rho _2\equiv \rho_{\rm Wigner} (x)$  back into the integral equations (\ref{inteq1}), (\ref{inteq2}). One gets a non-sensical result if $\lambda _1$, $\lambda _2$ are different. This is a rather disturbing feature of the leading-order solution that only relies on the sum of the two equations. To accommodate the difference, the solution needs to be refined.

Since $\mu _a\gg 1$, the kernels in the integral equations can be approximated by their large-distance asymptotics:
\begin{equation}
 K(x)\simeq x\ln x^2+2\gamma x+\frac{1}{6x}\equiv K^\infty (x).
\end{equation}
The Wigner distribution and its cousins have simple convolution with the asymptotic kernel:
\begin{eqnarray}
 &&\int_{-\mu }^{\mu }dy\,\sqrt{\mu ^2-y^2}\,K^\infty (x-y)
 =\frac{\pi }{3}\,x^3+ \left(\pi \mu ^2\ln\frac{\mu \,{\rm e}\,^{\gamma +\frac{1}{2}}}{2}+\frac{\pi }{6}\right)x
\nonumber \\
&&\int_{-\mu }^{\mu }dy\,\,\frac{K^\infty (x-y)}{\sqrt{\mu ^2-y^2}}=2\pi x\ln\frac{\mu \,{\rm e}\,^{\gamma +1}}{2}
\nonumber \\
&&
\int_{-\mu }^{\mu }dy\,\,
\frac{K^\infty (x-y)}{\left(\mu ^2-y^2\right)^{n+\frac{1}{2}}}
=-\frac{2^n(n-1)!\pi }{(2n-1)!!\mu ^{2n}}\,x,\qquad n=1,2,\ldots 
\end{eqnarray}
This observation suggests the following ansatz:
\begin{equation}\label{ansatz-bulk}
 \rho _a(x)=A\sqrt{\mu_a ^2-x^2}+\frac{2\mu _aAB_a}{\sqrt{\mu_a ^2-x^2}}
 +\frac{4\mu _a^2AC_a}{\left(\mu _a^2-x^2\right)^{\frac{3}{2}}}+\ldots 
\end{equation}
Each consecutive term adds an extra power of $1/\mu $, and hence of $1/ \sqrt{\lambda }$, so this ansatz naturally represents the strong-coupling expansion of the density. While  $\mu _1=\mu _2$ at the leading order, due to the locking effect, the two endpoints split at higher orders. On the contrary, the overall normalization constant $A$ must remain the same to all orders in $1/\sqrt{\lambda }$, as will become clear shortly.

The asymptotic integral operators generate only cubic and linear terms in $x$ at each order in $1/\mu $. Moreover, the cubic terms only arise from the Wigner function. Cancellation of the cubic terms is precisely the condition that the overall constant $A$ is the same for the two densities.
But the linear terms do not cancel automatically. Matching them gives two scalar equations:
\begin{eqnarray}\label{long-eq}
 &&1-\mu _{1,2}^2\ln\frac{\mu _{1,2}\,{\rm e}\,^{\gamma +\frac{1}{2}}}{2}+\mu _{2,1}^2\ln\frac{\mu _{2,1}\,{\rm e}\,^{\gamma +\frac{1}{2}}}{2}
 -4B_{1,2}\mu _{1,2}\ln\frac{\mu _{1,2}\,{\rm e}\,^{\gamma +1}}{2}
 \nonumber \\ &&
+4B_{2,1}\mu _{2,1}\ln\frac{\mu _{2,1}\,{\rm e}\,^{\gamma +1}}{2}
+8C_{1,2}-8C_{2,1}=\frac{8\pi }{A\lambda _{1,2}}\,.
\end{eqnarray}

The unit normalization of the densities gives another two conditions that can be used to eliminate $B_a$:
\begin{equation}\label{Ba}
 B_a=\frac{1}{2\pi A\mu _a}-\frac{\mu _a}{4}\,.
\end{equation}
When (\ref{Ba}) is substituted in (\ref{long-eq}) the latter considerably simplifies:
\begin{equation}
 1+\frac{\mu _{1,2}^2}{2}-\frac{\mu _{2,1}^2}{2}-\frac{2}{\pi A}\,\ln\frac{\mu _{1,2}}{\mu _{2,1}}+8C_{1,2}-8C_{2,1}=\frac{8\pi }{A\lambda _{1,2}}\,.
\end{equation}
The sum of the two equations determines $A$:
\begin{equation}
 A=\frac{4\pi }{\lambda _1}+\frac{4\pi }{\lambda _2}=\frac{8\pi }{\lambda }\,,
\end{equation}
while their difference gives:
\begin{equation}\label{mainconstraint}
 \mu _1^2-\mu_2^2-\frac{\lambda }{2\pi ^2}\,\ln\frac{\mu _1}{\mu _2}+
 16(C_1-C_2)=\lambda\left(\frac{1}{\lambda _1}-\frac{1}{\lambda _2}\right).
\end{equation}

\begin{figure}[t]
\begin{center}
 \centerline{\includegraphics[width=10cm]{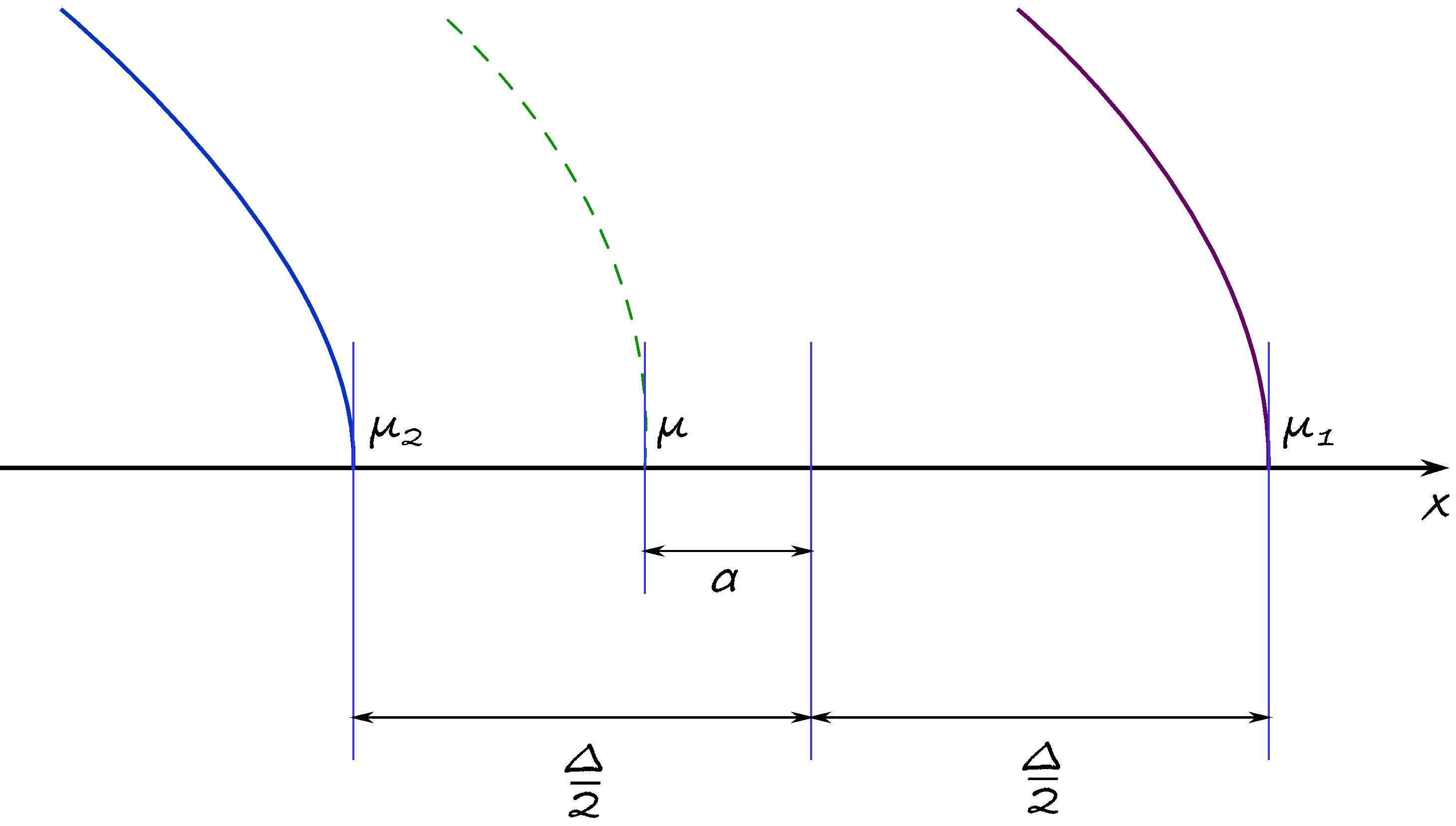}}
\caption{\label{edges}\small The endpoint structure of the eigenvalue distribution: $\Delta $ is the gap between $\mu _1$ and $\mu _2$, while $\alpha $ is the offset of the midpoint from the Gaussian-model prediction $\mu =\sqrt{\lambda }/2\pi $ (see also fig.~\ref{densities-fig}).}
\end{center}
\end{figure}

The constants $B_a$ should stay finite in the large-$\lambda $ limit, which requires cancellation between the two terms in (\ref{Ba}), nominally of order $\mathcal{\mathcal{O}}(\sqrt{\lambda })$ each. This requirement fixes $\mu _a=\sqrt{\lambda }/2\pi +\mathcal{O}(1)$. If we parameterize the endpoints of the eigenvalue distributions as in fig.~\ref{edges}:
\begin{equation}\label{muzy}
 \mu _{1,2}=\frac{\sqrt{\lambda }}{2\pi }+\alpha \pm\frac{\Delta }{2}\,,
\end{equation}
the normalization condition (\ref{Ba}) boils down to
\begin{equation}\label{B12}
 B_{1,2}=-\frac{\alpha }{2}\mp\frac{\Delta }{4}\,.
\end{equation}
All terms of order $\mathcal{O}(\lambda) $ in (\ref{mainconstraint}) also neatly cancel leaving behind one more equation:
\begin{equation}\label{C-condition}
 \alpha \Delta +4(C_1-C_2)=\frac{1}{2}-\frac{\theta }{2\pi }\,,
\end{equation}
with the $\theta $-parameter introduced in (\ref{theta}).

All in all, the saddle-point equations and normalization conditions fix $A$ and $B_a$ and impose one constraint on the four remaining variables, $\mu _{a}$ and $C_a$, or $\alpha $, $\Delta $ and $C_a$. It seems that the ansatz (\ref{ansatz-bulk}) introduces more unknowns than the equations can fix. At the same time, general theorems \cite{Gakhov} guarantee uniqueness of the solution to (\ref{inteq1}), (\ref{inteq2}). We found a three-parametric family. Why do general theorems fail? An apparent  contradiction is resolved if we recall that the general theorems rely on the boundary conditions at the endpoints in a crucial way \cite{Gakhov},  while the correct boundary behavior breaks down for the ansatz (\ref{ansatz-bulk}). The density explodes at the endpoints starting with the second order, allowing the ansatz to evade the uniqueness theorems. This also means that the ansatz is not applicable for $x$ very close to $\pm \mu _a$, and indeed at $x\pm \mu_a \sim \mathcal{O}(1)$ all the terms in the expansion are of the same order signaling the breakdown of the strong-coupling expansion. The equations have to be solved separately near the boundary. It will become clear later that matching to the bulk will eventually fix all the remaining ambiguities.

\section{Boundary}

The bulk solution suggests the following behavior near the endpoints:
\begin{equation}\label{scaling}
 \rho _a(x)\simeq A\sqrt{2\mu _a}\,f_a(\mu _a-x),
\end{equation}
where  $f_{1,2}(\xi )$ are some order-one scaling functions.
Their large-distance asymptotics is fixed by matching to the bulk solution (\ref{ansatz-bulk}):
\begin{equation}
 f_a (\xi )\stackrel{\xi \rightarrow \infty }{\simeq }\sqrt{\xi }+\frac{B_a}{\sqrt{\xi }}+\frac{C_a}{\xi ^{\frac{3}{2}}}\equiv f_a^\infty (\xi ).
\end{equation}

Integral equations for the scaling functions can be derived in two steps. The difficulty lies in the non-locality of the original, exact saddle-point equations. Even if we zoom in onto the spectral edge, the integrals would receive contributions from the whole eigenvalue interval. To isolate the boundary region we can use the following trick \cite{Chen:2014vka}. Consider exact saddle-point equations, schematically written as
\begin{equation}
 R_{ab}*\rho _b=\frac{8\pi ^2}{\lambda _a}\,x,
\end{equation}
where $*$ represents convolution. The perturbative bulk solution satisfies
\begin{equation}
  R^\infty _{ab}*\rho^\infty  _b=\frac{8\pi ^2}{\lambda _a}\,x,
\end{equation}
where $R^\infty $ is $R$  with $K$ replaced by $K^\infty $. This equation is actually exact, inspite of all approximations made. Hence,
\begin{equation}
 R*\rho= R^\infty *\rho^\infty.
\end{equation}
Subtracting $R*\rho ^\infty $ from both sides we get:
\begin{equation}
 R*(\rho -\rho ^\infty )=(R^\infty -R)*\rho ^\infty .
\end{equation}

These formal manipulations achieve our goal. Now taking $x=\mu -\xi $ with $\xi =\mathcal{O}(1)$, we find that only $y=\mu -\eta $ with $\eta =\mathcal{O}(1)$ contribute to the convolution integrals. Indeed, $R(\xi-\eta )$ grows as $(\eta -\xi )\ln(\eta -\xi) $, but $\rho -\rho ^\infty $ decays as $\eta ^{-5/2}$ away from the boundary. The convolution integral in $R*(\rho -\rho ^\infty )$ thus converges  and can be extended to infinity. Likewise, $\rho ^\infty $ grows as $\eta  ^{1/2}$, but $R-R^\infty $ decays as $1/(\eta -\xi) ^2$, so all integrals converge and the upper limit of integration can be safely removed:
\begin{equation}\label{boundaryinteq}
 \int_{0}^{\infty }R_{ab}(\xi -\eta )\left(f_b(\eta )-f^\infty _b(\eta )\right)
 =\int_{0}^{\infty }\left(R^\infty _{ab}(\xi -\eta )-R_{ab}(\xi -\eta )\right)f_b^\infty (\eta ).
\end{equation}

The explicit form of the kernel in the last equation is
\begin{equation}
 R_{ab}(\xi )=\begin{pmatrix}
 \frac{1}{\xi }-K(\xi)  & K(\xi -\Delta ) \\ 
 K(\xi +\Delta )  &  \frac{1}{\xi }-K(\xi)  \\ 
 \end{pmatrix},
\end{equation}
and the same for $R^\infty $ with $K\rightarrow K^\infty $. The shift by $\Delta $  in the off-diagonal terms occurs because of the gap between the endpoints of $\rho _1$ and $\rho _2$ (fig.~\ref{edges}) and the way we have defined the scaling functions in  (\ref{scaling}).

The resulting equation is of the Wiener-Hopf type and can be solved by Fourier transform
\begin{equation}
 f_a(\xi )=\int_{-\infty }^{+\infty }\frac{d\omega }{2\pi }\,\,\,{\rm e}\,^{-i\omega \xi }f_a(\omega ).
\end{equation}
Since $f_a(\xi )=0$ for $\xi <0$, its Fourier image is analytic in the upper half plane of $\omega $. 

The integral equation cannot be straightforwardly Fourier transformed, because it holds only for positive $\xi $. The equation can be extended to the whole real line at the expense of introducing another unknown function,  different from zero at negative $\xi $. After that the equation can be integrated and becomes algebraic in the Fourier space:
\begin{equation}\label{WH-bare}
 R(f-f^\infty )=(R^\infty -R)f^\infty +X_-.
\end{equation}
The subscript indicates that $X_-$ vanishes for $\xi >0$ and is therefore negative-half-plane analytic function of $\omega $.

The Wiener-Hopf method is based on the analytic factorization of the kernel:
\begin{equation}\label{Riemann-Hilbert}
 G_-R=G_+,
\end{equation}
where $G_\pm$ are matrix functions analytic in the upper/lower half-planes. Multiplying the two sides of (\ref{WH-bare}) by $G_-$, we get:
\begin{equation}\label{WH-dressed}
 G_+(f-f^\infty )=(G_-R^\infty -G_+)f^\infty +G_-X_-.
\end{equation}
This equation contains two unknown functions, $f$ and $X_-$, but they are analytic in different halves of the complex plane and can be disentangled with the help of the projection operators:
\begin{equation}\label{pm-projector}
 \mathcal{F}_\pm(\omega )=\pm\int_{-\infty }^{+\infty }\frac{d\nu }{2\pi i}\,\,
 \frac{\mathcal{F}(\nu )}{\nu -\omega \mp i\epsilon }\,,
\end{equation}
that singles out a half-plane analytic part of  $\mathcal{F}$.

The $+$ projection of (\ref{WH-dressed}) gives:
\begin{equation}
 G_+(f-f^\infty )=\left[(G_-R^\infty -G_+)f^\infty \right]_+.
\end{equation}
Linearity of the projection and upper-half-plane analyticity of $f^\infty $ then give:
\begin{equation}
 f=G_+^{-1}\left[G_-R^\infty f^\infty \right]_+.
\end{equation}
This equation constitutes a formal solution of the boundary problem. It still remains to analytically factorize the kernel.

The Fourier images of the functions appearing in the construction are
\begin{eqnarray}\label{Rw}
 R(\omega )&=&2\pi i \mathop{\mathrm{sign}}\omega \coth\frac{\omega }{2}\begin{bmatrix}
 \coth \omega   & -\frac{\,{\rm e}\,^{i\Delta \omega }}{\sinh\omega }  \\ 
   -\frac{\,{\rm e}\,^{-i\Delta \omega }}{\sinh\omega }   & \coth\omega  \\ 
 \end{bmatrix}
 \\
 R^\infty (\omega )&=&\frac{4\pi i\mathop{\mathrm{sign}}\omega }{\omega ^2}\begin{bmatrix}
 1+\frac{5\omega ^2}{12}   & \left(-1+\frac{\omega ^2}{12}\right)\,{\rm e}\,^{i\Delta \omega }  \\ 
    \left(-1+\frac{\omega ^2}{12}\right)\,{\rm e}\,^{-i\Delta \omega } & 1+\frac{5\omega ^2}{12}  \\ 
 \end{bmatrix}
 \\
 \label{fa-inf}
 f^\infty_a (\omega )&=&\frac{\sqrt{\pi }\,i^{\frac{3}{2}}}{2(\omega +i\epsilon )^{\frac{3}{2}}}\left(1-2i\omega B_a+4\omega ^2C_a\right).
\end{eqnarray}
The analytic form of $\mathop{\mathrm{sign}}\omega $ is implied here:
\begin{equation}
 \mathop{\mathrm{sign}}\omega =\lim_{\epsilon \rightarrow 0}
 \frac{\sqrt{\omega +i\epsilon }}{\sqrt{\omega -i\epsilon }}\,,
\end{equation}
where the branch cut of $\sqrt{\omega \mp i\epsilon }$ extends into the upper/lower half-plane. 

Incidentally, the fractional powers of $\omega +i\epsilon $ cancel in the product $R^\infty f^\infty $, leaving  a triple pole $\omega =-i\epsilon $ as the only singularity in the lower half-plane. Closing the contour of the $+$ projection in the lower half-plane picks the residue:
\begin{equation}\label{residue-formula}
 f(\omega )=G^{-1}_+(\omega )\mathop{\mathrm{res}}_{\nu =0}
 \frac{G_-(\nu )R^\infty (\nu )f^\infty (\nu )}{\omega -\nu }\,.
\end{equation}
This equation expresses the scaling functions $f_a$ through the Wiener-Hopf factors of the kernel. The problem reduces to analytic factorization of the matrix function (\ref{Rw}) according to (\ref{Riemann-Hilbert}).

Analytic matrix factorization is known as the Riemann-Hilbert problem and has numerous applications in the theory of solitons \cite{Faddeev:1987ph} and in algebraic geometry. For a scalar function ($1\times 1$ matrix), the problem can be solved in quadratures by taking the logarithm, applying the projection (\ref{pm-projector}) and exponentiating back. This procedure does not work for matrices due to non-commutativity of matrix multiplication. Matrix factorization is a substantially more complicated problem (see \cite{Its-RH-review} for a review) for which there is no simple plug-in solution. Fortunately, for the particular case of (\ref{Rw})  the Riemann-Hilbert factorization  has been carried out explicitly \cite{Antipov}. The Wiener-Hopf factors were found in   \cite{Antipov} by exploiting analytic properties of the hypergeometric functions and linear identities among them. In principle, an explicit formula is all we need, but we would like to present a derivation that highlights connections to the inverse scattering problem. This perspective can be useful in view of possible generalizations and may hint on the links to integrability of the dual string theory \cite{Bena:2003wd,Kazakov:2004qf}.

\subsection{Matrix factorization}

Consider Schr\"odinger equation with the P\"oschl-Teller potential:
\begin{equation}
 -\frac{d^2\psi }{dx^2}+\frac{1}{4\cosh^2x}\,\psi =k^2\psi .
\end{equation}
Its scattering theory is conveniently formulated in terms of the Jost functions characterized by purely exponential asymptotics at infinity:
\begin{equation}
 \psi _L^\pm\simeq \,{\rm e}\,^{\mp ikx}~~(x\rightarrow -\infty ),\qquad 
 \psi _R^\pm\simeq \,{\rm e}\,^{\pm ikx}~~(x\rightarrow +\infty ).
\end{equation}
The Jost functions  $\psi ^-_{L,R}$ describe in-type scattering states with the incident wave moving left or right and the amplitude of the transmitted wave normalized to one, while $\psi _{L,R}^+$ are the $T$-conjugate out-states. The four Jost functions are related by parity and complex conjugation.

The Jost functions admit analytic continuation into the complex momentum plane. Moreover, $\psi _{L,R}^+$ are analytic in the upper half-plane and $\psi _{L,R}^-$ are analytic in the lower half-plane, after oscillating exponentials are knocked off:
\begin{equation}
 \chi _{L,R}^+=\,{\rm e}\,^{\pm ikx}\psi^+ _{L,R},\qquad 
 \chi _{L,R}^-=\,{\rm e}\,^{\mp ikx}\psi^- _{L,R}.
\end{equation}
These functions are faithfully half-plane analytic in $k$.

For the P\"oschl-Teller potential the Jost functions can be found explicitly:
\begin{eqnarray}
 \psi _R^\pm&=&\,{\rm e}\,^{\pm ikx}\sqrt{1+\,{\rm e}\,^{-2x}}\,{}_2\!F_1\left(\frac{1}{2}\mp ik,\frac{1}{2}\,;1\mp ik;-\,{\rm e}\,^{-2x}\right)
\nonumber \\
\psi _L^\pm&=&\,{\rm e}\,^{\mp ikx}\sqrt{1+\,{\rm e}\,^{2x}}\,{}_2\!F_1\left(\frac{1}{2}\mp ik,\frac{1}{2}\,;1\mp ik;-\,{\rm e}\,^{2x}\right).
\nonumber 
\end{eqnarray}
The four Jost functions are linearly dependent, because they are solutions of a second-order differential equation, and all of them can be expressed through any two chosen as the basis. 

In the case at hand, the linear relations follow from transformation rules of the hypergeometric function under argument inversion. For example, applying the $x\rightarrow -x$ transformation to $\psi _R^\pm$, we get:
\begin{equation}\label{left-right}
 \psi _R^\pm=\mp\frac{i}{\sinh\pi k}\,\psi _L^\pm\pm i\coth\pi k\,
 \frac{B\left(\frac{1}{2}\pm ik,\frac{1}{2}\right)}{B\left(\frac{1}{2}\mp ik,\frac{1}{2}\right)}\,
 \psi _L^\mp.
\end{equation}

More conventionally, the in-states are chosen as the basis. The out-states are then related to them by the S-matrix. Reshuffling (\ref{left-right}) we find:
\begin{equation}
 \begin{bmatrix}
  \psi ^+_L & \psi ^+_R \\ 
 \end{bmatrix}
 = i\,\frac{B\left(\frac{1}{2}+ ik,\frac{1}{2}\right)}{B\left(\frac{1}{2}- ik,\frac{1}{2}\right)}\,
 \begin{bmatrix}
  \psi ^-_R & \psi ^-_L \\ 
 \end{bmatrix}
 \begin{bmatrix}
 \tanh \pi k  & -\frac{i}{\cosh\pi k} \\ 
   -\frac{i}{\cosh\pi k}  &  \tanh \pi k \\ 
 \end{bmatrix}.
\end{equation}
The same relation holds for the derivatives of the Jost functions and hence for their Wronskians
\begin{equation}
 W^+=\begin{bmatrix}
  \psi _L^+ & \psi _R^+ \\ 
 \frac{d\psi _L^+}{dx}  & \frac{d\psi _R^+}{dx} \\ 
 \end{bmatrix},\qquad 
  W^-=\begin{bmatrix}
  \psi _R^- & \psi _L^- \\ 
 \frac{d\psi _R^-}{dx}  & \frac{d\psi _L^-}{dx} \\ 
 \end{bmatrix}.
\end{equation}
Namely,
\begin{equation}
 W^+=W^-S.
\end{equation}
This is already close to what we need. One can say that Wronskians factorize the S-matrix, but Wronskians by themselves are not yet analytic. The oscillating factors in the Jost functions have to be offset by a similarity transformation:
\begin{equation}
 W^\pm\rightarrow W^\pm\Omega ,\qquad S\rightarrow \Omega^{-1} S\Omega
\end{equation}
with 
$$
\Omega =\mathop{\mathrm{diag}}(\,{\rm e}\,^{ikx+\frac{x}{2}},\,{\rm e}\,^{-ikx-\frac{x}{2}}).
$$

The truly analytic factorization formula is slightly more complicated:
\begin{eqnarray}
&& B\left(\frac{1}{2}- ik,\frac{1}{2}\right)
 \begin{bmatrix}
  \psi _L^+\,{\rm e}\,^{ikx+\frac{x}{2}} &  \psi _R^+ \,{\rm e}\,^{-ikx-\frac{x}{2}} 
   \\ 
 \frac{d\psi _L^+}{dx}\,\,{\rm e}\,^{ikx+\frac{x}{2}}   &   
   \frac{d\psi _R^+}{dx}\,\,{\rm e}\,^{-ikx-\frac{x}{2}} \\
 \end{bmatrix}
\nonumber \\
\nonumber
&&=iB\left(\frac{1}{2}+ ik,\frac{1}{2}\right)
    \begin{bmatrix}
  \psi _R^-\,{\rm e}\,^{ikx+\frac{x}{2}} &  \psi _L^- \,{\rm e}\,^{-ikx-\frac{x}{2}} 
   \\ 
 \frac{d\psi _R^-}{dx}\,\,{\rm e}\,^{ikx+\frac{x}{2}}   &   
   \frac{d\psi _L^-}{dx}\,\,{\rm e}\,^{-ikx-\frac{x}{2}} \\
 \end{bmatrix}
    \begin{bmatrix}
 \tanh \pi k  & -\frac{i\,{\rm e}\,^{-2ikx-x}}{\cosh\pi k} \\ 
   -\frac{i\,{\rm e}\,^{2ikx+x}}{\cosh\pi k}  &  \tanh \pi k \\ 
 \end{bmatrix}.
\end{eqnarray}
Remarkably, the similarity transformation not only rendered all wavefunction half-plane analytic, but also brought the S-matrix into the form very similar to (\ref{Rw}). In fact, $\Omega^{-1} S\Omega$  coincides with $R(\omega )$  up to an overall scalar factor after the following change of variables:
\begin{equation}
 k\rightarrow \frac{\omega }{\pi }+\frac{i}{2}\,,\qquad 
 x\rightarrow -\frac{\pi \Delta }{2}\,.
\end{equation}
The scalar factor is easily factorizable by itself:
\begin{equation}
 \frac{1}{2\pi ^2}\,\mathop{\mathrm{sign}}\omega \coth\frac{\omega }{2}=\frac{1}{\sqrt{\omega +i\epsilon }\,B\left(\frac{1}{2}-\frac{i\omega }{2\pi }\,,\frac{1}{2}\right)}\,
 \cdot \,
 \frac{1}{\sqrt{\omega -i\epsilon }\,B\left(\frac{1}{2}+\frac{i\omega }{2\pi }\,,\frac{1}{2}\right)}\,.
\end{equation}
The solution of the Riemann-Hilbert problem thus follows from the scattering theory of the P\"oschl-Teller potential!

The final result is rather bulky, and is best written in the shorthand notation:
\begin{equation}\label{Q-function}
 Q(\alpha ,\beta ;q)=B(\alpha ,\beta ){}_2\!F_1(\alpha ,\beta ;\alpha +\beta ;-q).
\end{equation}
The salient properties of this function are summarized in the appendix. The solution of the Riemann-Hilbert problem (\ref{Riemann-Hilbert}) takes the following form: 
\begin{eqnarray}
 G_+&=&\frac{4\pi ^2}{\sqrt{\omega +i\epsilon }\,\,B\left(\frac{1}{2}-\frac{i\omega }{2\pi }\,,\frac{1}{2}\right)}
\begin{bmatrix}
   a_+ & b_+ \\ 
  c_+ & d_+  \\ 
 \end{bmatrix}
 \begin{bmatrix}
  \,{\rm e}\,^{-\frac{\pi \Delta }{2}} & 0 \\ 
 0  &  \,{\rm e}\,^{\frac{\pi \Delta }{2}} \\ 
 \end{bmatrix}
 \\
 G_-&=&\frac{1}{\pi }\,\sqrt{\omega -i\epsilon }\,B\left(\frac{1}{2}+\frac{i\omega }{2\pi }\,,\frac{1}{2}\right)
 \begin{bmatrix}
  a_- & b_- \\ 
  c_- & d_-  \\ 
 \end{bmatrix}
\end{eqnarray}
with
\begin{eqnarray}
 a_+&=&Q\left(1-\frac{i\omega }{\pi }\,,\frac{1}{2}\,;\,{\rm e}\,^{-\pi \Delta }\right)
\nonumber \\
b_+&=&Q\left(1-\frac{i\omega }{\pi }\,,\frac{1}{2}\,;\,{\rm e}\,^{\pi \Delta }\right) 
\nonumber \\
c_+&=&\left(\frac{1}{2}-\frac{i\omega }{\pi }\right)Q\left(1-\frac{i\omega }{\pi }\,,\,\frac{1}{2}\,;\,{\rm e}\,^{-\pi \Delta }\right)+\frac{1}{1+\,{\rm e}\,^{\pi \Delta }}\,Q\left(1-\frac{i\omega }{\pi }\,,\,\frac{3}{2}\,;\,{\rm e}\,^{-\pi \Delta }\right) 
\nonumber \\
d_+&=&-\left(\frac{1}{2}-\frac{i\omega }{\pi }\right)Q\left(1-\frac{i\omega }{\pi }\,,\,\frac{1}{2}\,;\,{\rm e}\,^{\pi \Delta }\right)-\frac{1}{1+\,{\rm e}\,^{-\pi \Delta }}\,Q\left(1-\frac{i\omega }{\pi }\,,\,\frac{3}{2}\,;\,{\rm e}\,^{\pi \Delta }\right) 
\nonumber \\
a_-&=&Q\left(\frac{i\omega }{\pi }\,,\,\frac{1}{2}\,;\,{\rm e}\,^{\pi \Delta }\right)
\nonumber \\
b_-&=&Q\left(\frac{i\omega }{\pi }\,,\,\frac{1}{2}\,;\,{\rm e}\,^{-\pi \Delta }\right)
\nonumber \\
c_-&=&\left(\frac{1}{2}-\frac{i\omega }{\pi }\right)Q\left(\frac{i\omega }{\pi }\,,\,\frac{1}{2}\,;\,\,{\rm e}\,^{\pi \Delta }\right)-\frac{1}{1+\,{\rm e}\,^{-\pi \Delta }}\,Q\left(\frac{i\omega }{\pi }\,,\,\frac{3}{2}\,;\,{\rm e}\,^{\pi \Delta }\right)
\nonumber \\
d_-&=&-\left(\frac{1}{2}-\frac{i\omega }{\pi }\right)Q\left(\frac{i\omega }{\pi }\,,\,\frac{1}{2}\,;\,\,{\rm e}\,^{-\pi \Delta }\right)+\frac{1}{1+\,{\rm e}\,^{\pi \Delta }}\,Q\left(\frac{i\omega }{\pi }\,,\,\frac{3}{2}\,;\,{\rm e}\,^{-\pi \Delta }\right).
\end{eqnarray}

We also need the inverse of $G_+$. The standard Wronskian identity appears useful in that regard:
\begin{equation}
 \begin{bmatrix}
 \psi _L^+  & \psi _R^+ \\ 
  \frac{d\psi _L^+}{dx} & \frac{d\psi _R^+}{dx} \\ 
 \end{bmatrix}^{-1}=\frac{1}{2ik}
 \begin{bmatrix}
  \frac{d\psi^+ _R}{dx} & -\psi _R^+ \\ 
  -\frac{d\psi _L^+}{dx} & \psi _L^+ \\ 
 \end{bmatrix}.
\end{equation}
Using this identity we get:
\begin{equation}
 G^{-1}_+=\frac{\sqrt{\omega +i\epsilon }}{8\pi ^3}\,B\left(\frac{1}{2}-\frac{i\omega }{2\pi }\,,\frac{1}{2}\right)
 \begin{bmatrix}
  1+\,{\rm e}\,^{\pi \Delta } & 0 \\ 
   0 &  1+\,{\rm e}\,^{-\pi \Delta } \\ 
 \end{bmatrix}
 \begin{bmatrix}
-d_+  & b_+  \\ 
  c_+ & -a_+ \\ 
 \end{bmatrix}
\end{equation}
Checking that $G_+^{-1}G_+=1$ by a direct calculation is a really fun exercise.

\subsection{Solving the boundary problem}

With all the ingredients at hand, we can now find the scaling functions from (\ref{residue-formula}).  Evaluating the residue with the help of (\ref{small-alpha-Q}) we get:
\begin{equation}\label{prelim-sol}
 f(\omega )=\frac{2\pi ^{\frac{5}{2}}i^{\frac{3}{2}}}{\omega ^2}\,G_+^{-1}(\omega )\left(u+\frac{i\omega }{\pi }\,v\right),
\end{equation}
with
\begin{eqnarray}
 u&=&\begin{bmatrix}
 1  \\ 
  -\frac{1}{2}\,\tanh\frac{\pi \Delta }{2} \\ 
 \end{bmatrix}
\nonumber \\
\label{v-vec}
v&=&\left(\pi \alpha-\ln\cosh\frac{\pi \Delta }{2}\right)
\begin{bmatrix}
 1  \\ 
  -\frac{1}{2}\,\tanh\frac{\pi \Delta }{2} \\ 
 \end{bmatrix}
+\begin{bmatrix}
 0  \\ 
 \tanh\frac{\pi \Delta }{2}  \\ 
 \end{bmatrix},
\end{eqnarray}
where the explicit form of $B_{1,2}$ from (\ref{B12}) has been used. 

The densities should vanish as a square root at the boundary and so should the scaling functions $f_a(\xi )$. The right behavior at $\xi =0$ is not at all guaranteed for the solution obtained above and has to be imposed by hand as an extra condition. 
The endpoint behavior in the coordinate space is determined by the dependence of the Fourier image on large imaginary frequencies. The square root maps to $\omega ^{-3/2}$ in the Fourier space, and the right boundary conditions correspond to
\begin{equation}
 f_a(i\pi \kappa )\stackrel{\kappa \rightarrow +\infty }{\simeq }
 \frac{Z_a}{\kappa ^\frac{3}{2}}
\end{equation}
with some constant $Z_a$.

The general solution as given above is not consistent with this requirement. An  expansion of $G_+^{-1}$ at large imaginary frequencies follows from (\ref{large-alpha-Q}), and starts with $\kappa ^{1/2}$:
\begin{equation}
 G^{-1}_+(i\pi \kappa )\stackrel{\kappa \rightarrow +\infty }{=}
 \frac{\sqrt{i\kappa }}{4\sqrt{2}\,\pi ^2}\begin{bmatrix}
  \sqrt{1+\,{\rm e}\,^{\pi \Delta }} & 0 \\ 
  \sqrt{1+\,{\rm e}\,^{-\pi \Delta }}  &  0 \\ 
 \end{bmatrix}+\mathcal{O}\left(\frac{1}{\sqrt{\kappa }}\right),
\end{equation}
which means that in general $f(i\pi \kappa )$ will scale as $\kappa ^{-1/2}$ because of the $v$-term in (\ref{prelim-sol}). In the coordinate space $1/\sqrt{\kappa }$ translates to $1/\sqrt{\xi }$, an expected asymptotics of a generic solution to the integral equation \cite{Gakhov}. But we are seeking a special solution where this leading asymptotic cancels leaving behind the desired $\sqrt{\xi }$  behavior.
This happens  if
\begin{equation}
 \begin{bmatrix}
  \sqrt{1+\,{\rm e}\,^{\pi \Delta }} & 0 \\ 
  \sqrt{1+\,{\rm e}\,^{-\pi \Delta }}  &  0 \\ 
 \end{bmatrix}v=0.
\end{equation}
The next term scales as $\kappa ^{-3/2}$ and if this condition is imposed the solution has the right boundary asymptotics.

One may expect that the boundary conditions impose two constraints for each of the two independent functions,
 but $G_+^{-1}(i\pi \kappa )$ degenerates as a matrix at $\kappa \rightarrow +\infty $ and, as a result, only one condition survives. The condition is actually very simple, it basically requires the top component of $v$ to vanish. 
From the explicit formula (\ref{v-vec}) we find that this is equivalent to
\begin{equation}\label{alpha-delta}
 \alpha =\frac{1}{\pi }\,\ln\cosh\frac{\pi \Delta }{2}\,.
\end{equation}
We get an extra constraint, invisible in the bulk, that relates two of the remaining four parameters of the solution. 

Interestingly, $\alpha $ appears to be always positive. This implies the following inequality:
\begin{equation}
 \frac{\mu _1+\mu _2}{2}\geq \mu ,
\end{equation}
illustrated in fig.~\ref{edges}. The density for the weaker coupling ($\rho _2$) squeezes compared to the Wigner semicircle, while the density for the larger coupling ($\rho _1$) expands. This is intuitively clear, because the extent of the density is controlled by the overall linear force inversely proportional to the coupling. What is less obvious is that the expansion of $\rho _1$ is always more pronounced than the squeezing of $\rho _2$. It would be interesting to understand this behavior at a qualitative level.

The boundary solution really simplifies once the condition (\ref{alpha-delta}) is imposed. The scaling functions (\ref{prelim-sol})  become
\begin{eqnarray}
 f_{1,2}(\omega )&=&\frac{i^{\frac{3}{2}}B\left(\frac{1}{2}-\frac{i\omega }{2\pi }\,,\,\frac{1}{2}\right)}{4\sqrt{\pi }\,(\omega+i\epsilon ) ^{\frac{3}{2}}}
 \left[
 \left(1-\frac{2i\omega }{\pi }\right)Q\left(1-\frac{i\omega }{\pi }\,,\,\frac{1}{2}\,;\,{\rm e}\,^{\pm \pi \Delta }\right)
 \right.
\nonumber \\
&&\vphantom{\frac{i^{\frac{3}{2}}B\left(\frac{1}{2}-\frac{i\omega }{2\pi }\,,\,\frac{1}{2}\right)}{4\sqrt{\pi }\,(\omega+i\epsilon ) ^{\frac{3}{2}}}}
\left.
 +\,{\rm e}\,^{\pm \pi \Delta }
 Q\left(1-\frac{i\omega }{\pi }\,,\,\frac{3}{2}\,;\,{\rm e}\,^{\pm \pi \Delta }\right)
 \right].
\end{eqnarray}
They admit an integral representation
\begin{equation}\label{f12-omega}
 f_{1,2}(\omega )=\frac{i^{\frac{3}{2}}B\left(\frac{1}{2}-\frac{i\omega }{2\pi }\,,\,\frac{1}{2}\right)}{2\sqrt{\pi }\,(\omega+i\epsilon ) ^{\frac{3}{2}}}
 \int_{0}^{1}du\,\left(\frac{1+\,{\rm e}\,^{\pm\pi \Delta }u^2}{1-u^2}\right)^{\frac{i\omega }{\pi }}
 \left(1-\frac{2i\omega }{\pi }\,\,\frac{1}{1+\,{\rm e}\,^{\pm\pi \Delta }u^2}\right),
\end{equation}
that follows from  (\ref{integral-Q})  upon a change of variables $t=u^2$.
This form is particularly convenient for Taylor expansion at small $\omega $.

The scaling functions should match with the bulk solution at large $\xi $. In practice, matching means that the Taylor expansion at small $\omega $ coincides with (\ref{fa-inf}). The first three orders can be easily found from the integral representation:
\begin{eqnarray}
 f_{1,2}(\omega )&\stackrel{\omega \rightarrow 0}{=}&
 \frac{\sqrt{\pi }\,i^{\frac{3}{2}}}{2\omega ^{\frac{3}{2}}}\left[
 1+\frac{i\omega }{\pi }\ln\frac{1+\,{\rm e}\,^{\pm\pi \Delta }}{2} 
 \right.
\nonumber \\
&&\left.
 +\frac{\omega ^2}{\pi ^2}\left(
 \frac{\pi ^2}{8}-2\arctan^2\,{\rm e}\,^{\pm\frac{\pi \Delta }{2}}
 -\frac{1}{2}\,\ln^2\frac{1+\,{\rm e}\,^{\pm\pi \Delta }}{2}
 \right)+\ldots \right].
\end{eqnarray}
Comparing to (\ref{fa-inf}) we find that
\begin{equation}
 B_{1,2}=-\frac{1}{2\pi }\,\ln\cosh\frac{\pi \Delta }{2}\pm\frac{\Delta }{4}\,.
\end{equation}
Taking into account (\ref{alpha-delta}), this gives the same expression (\ref{B12}) that was inferred from the bulk normalization condition. We get nothing new, this is not even a consistency check because the first two orders are guaranteed to match by construction. 

New data is contained in the next term. Reading off its coefficient and comparing to (\ref{fa-inf}) we find:
\begin{equation}\label{C12}
 C_{1,2}=\frac{1}{32}-\frac{1}{2\pi ^2}\,\arctan^2\,{\rm e}\,^{\pm\frac{\pi \Delta }{2}}-\frac{1}{8\pi ^2}\left(\ln\cosh\frac{\pi \Delta }{2}\pm\frac{\pi \Delta }{2}\right)^2.
\end{equation}
This determines the two remaining unknowns and fixes all the parameters of the bulk solution.

\section{Wilson loops}

We can now complete the circle and use the remaining bulk condition (\ref{C-condition}) to find $\Delta $. To this end, we infer from (\ref{C12}) that
$$
 C_1-C_2=\frac{1}{8}-\frac{1}{2\pi }\,\arctan\,{\rm e}\,^{\frac{\pi \Delta }{2}}
 -\frac{\Delta }{4\pi }\,\ln\cosh\frac{\pi \Delta }{2}\,.
$$
Upon substitution of this formula along with (\ref{alpha-delta}) into  (\ref{C-condition}) many terms cancel, the relationship between $\Delta $ and $\theta $ simplifies and can be inverted, and at the end we find a simple analytic expression
\begin{equation}\label{Delta-theta}
 \Delta =\frac{2}{\pi }\,\ln\tan\frac{\theta }{4}\,.
\end{equation}
The other parameter that characterizes the eigenvalue distribution, $\alpha $, can be found from (\ref{alpha-delta}):
\begin{equation}
 \alpha =-\frac{1}{\pi }\,\ln\sin\frac{\theta }{2}\,.
\end{equation}

\begin{figure}[t]
\begin{center}
 \centerline{\includegraphics[width=8cm]{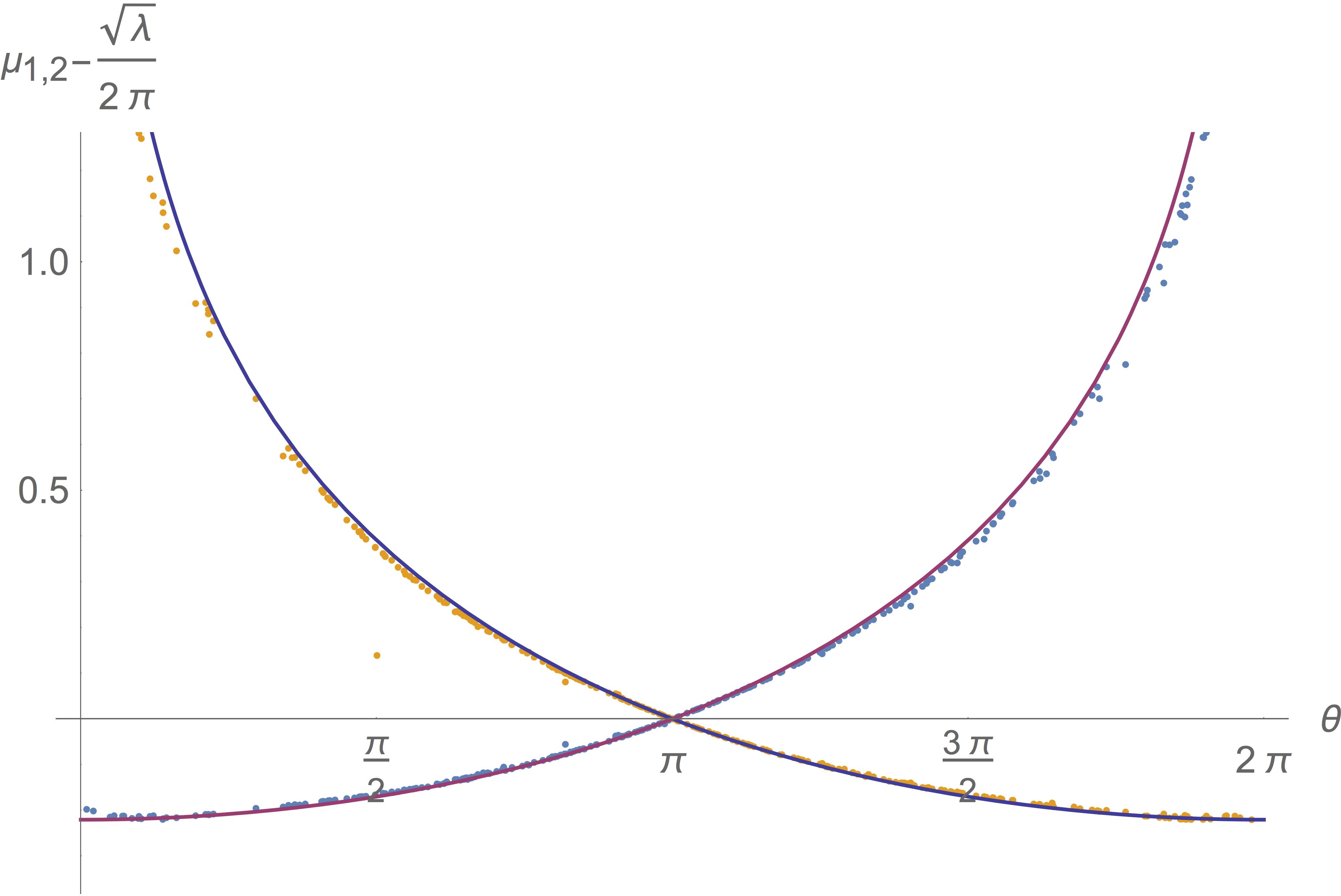}}
\caption{\label{pictorial-muz}\small The endpoint positions relative to the mean-field value $\mu =\sqrt{\lambda }/2\pi $, as functions of the $\theta $-parameter. The dots are obtained by picking $\lambda _1$, $\lambda _2$ randomly between $0$ and $8000$ and numerically solving the integral equations. Certain scatter in the numerical data is due to unaccounted $1/\sqrt{\lambda }$ corrections which are different for different points.}
\end{center}
\end{figure}

The endpoints are determined by the definition (\ref{muzy}):
\begin{eqnarray}
 \mu _1&=&\frac{\sqrt{\lambda }}{2\pi }-\frac{1}{\pi }\ln\left(2\cos^2\frac{\theta }{4}\right)+\mathcal{O}\left(\frac{1}{\sqrt{\lambda }}\right)
\nonumber \\
\mu _2&=&\frac{\sqrt{\lambda }}{2\pi }-\frac{1}{\pi }\ln\left(2\sin^2\frac{\theta }{4}\right)+\mathcal{O}\left(\frac{1}{\sqrt{\lambda }}\right)
\end{eqnarray}
This result is plotted in fig.~\ref{pictorial-muz}. The picture is symmetric under $\theta \rightarrow 2\pi -\theta $, $\mu _1\leftrightarrow \mu _2$, as expected.

The main contribution to the Wilson loop average (\ref{wils-int}) comes from the largest eigenvalues located near the edge of the distribution. The density under the integral in  (\ref{wils-int}) can thus be replaced by its scaling form (\ref{scaling}). Since the exponential weight guarantees fast convergence, the integration can be safely extended to infinity:
\begin{equation}
 W_a\simeq A\sqrt{2\mu _a}\,\,{\rm e}\,^{2\pi \mu _a}\int_{0}^{\infty }
 d\xi \,f_a(\xi )\,{\rm e}\,^{-2\pi \xi }.
\end{equation}
 The integral is the Fourier image of the scaling function at pure imaginary frequency:
\begin{equation}
 W_{1,2}\simeq 8\sqrt{\pi }\lambda ^{-\frac{3}{4}}\,{\rm e}\,^{\sqrt{\lambda }+2\pi \alpha \pm \pi \Delta }f_{1,2}(2\pi i).
\end{equation}

Using the explicit solution (\ref{f12-omega}) and substituting (\ref{alpha-delta}) for $\alpha $ we find:
\begin{equation}
 W_{1,2}=\cosh^2\frac{\pi \Delta }{2}\,
 \left(1\pm 2\sin\frac{\pi \Delta }{2}\,\arctan\,{\rm e}\,^{\pm\frac{\pi \Delta }{2}}
 \right)\sqrt{\frac{2}{\pi }}\,\lambda ^{-\frac{3}{4}}\,{\rm e}\,^{\sqrt{\lambda }}.
\end{equation}
Finally, expressing $\Delta $ as a function of $\theta $ with the help of (\ref{Delta-theta}), we obtain 
\begin{equation}\label{W1W2}
 W_1=  w(\theta )\sqrt{\frac{2}{\pi }}\,\lambda ^{-\frac{3}{4}}\,{\rm e}\,^{\sqrt{\lambda }},\qquad 
 W_2= w(2\pi -\theta )\sqrt{\frac{2}{\pi }}\,\lambda ^{-\frac{3}{4}}\,{\rm e}\,^{\sqrt{\lambda }},
\end{equation}
where
\begin{equation}
 w(\theta )=\frac{1-\frac{\theta }{2}\,\cot\frac{\theta }{2}}{\sin^2\frac{\theta }{2}}\,.
\end{equation}
This is the main result of the paper. 

\begin{figure}[t]
\begin{center}
 \centerline{\includegraphics[width=8cm]{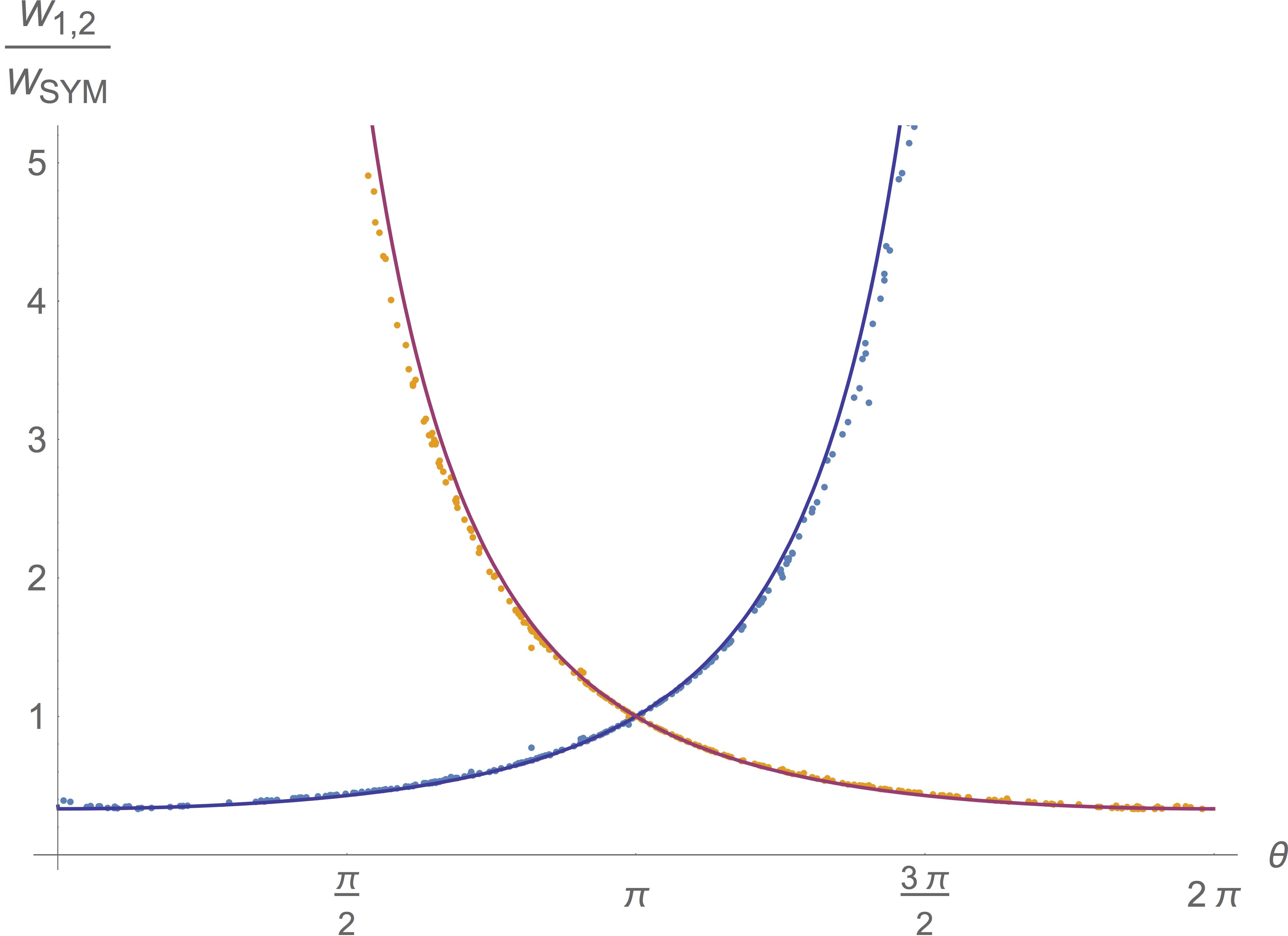}}
\caption{\label{wilson-figure}\small The circular Wilson loops in the quiver theory normalized by that in the $\mathcal{N}=4$ SYM, plotted as a function of the $\theta $-parameter. The dots represent the same data as in fig.~\ref{pictorial-muz}.}
\end{center}
\end{figure}

The function $w(\theta )$, shown in fig.~\ref{wilson-figure}, encodes the difference between the quiver CFT and $\mathcal{N}=4$  SYM. Indeed, the asymptotic strong-coupling expectation value in the SYM is given by (\ref{N=4WL}). Comparing to (\ref{W1W2}) we see that  $w(\theta )$ is an extra factor that arises in the quiver theory:
\begin{equation}
 \lim_{\lambda \rightarrow \infty }\frac{W_1}{W_{\rm SYM}}=w(\theta ),
 \qquad
 \lim_{\lambda \rightarrow \infty }\frac{W_2}{W_{\rm SYM}}=w(2\pi -\theta ). 
\end{equation}
 The ratio of Wilson loops is much easier to compute in string theory than a separate Wilson loop on its own. The disc amplitude for the circular loop in $AdS_5\times S^5$ has been known for a long time \cite{Drukker:2000ep} as a formal ratio of potentially divergent determinants. But in the ratio all divergences cancel making the Wilson loop normalized by its SYM counterpart an ideal playground for studying quantum string effects in holography  \cite{Forini:2015bgo,Faraggi:2016ekd}.

Observables better suited for comparison to string theory are the twisted and untwisted loop correlators:
\begin{equation}\label{wpm}
 w_\pm=\frac{W_1\pm W_2}{2W_{\rm SYM}}\,.
\end{equation}
The disc amplitude, normalized by the undeformed $AdS_5\times S^5$ counterpart, maps directly to $w_+$, while $w_-$ describes the disc with the twist operator inserted. Localization gives the following predictions at strong coupling:
\begin{equation}
 w_+(\theta )=\frac{1+\frac{\pi -\theta }{2}\cot\frac{\theta }{2}}{\sin^2\frac{\theta }{2}}\,,\qquad 
 w_-(\theta )=- \frac{\pi }{2}\,\,\frac{\cos\frac{\theta }{2}}{\sin^3\frac{\theta }{2}}\,.
\end{equation}
It would be very interesting to test these predictions by an explicit string-theory calculation.

\begin{figure}[t]
\begin{center}
 \centerline{\includegraphics[width=7cm]{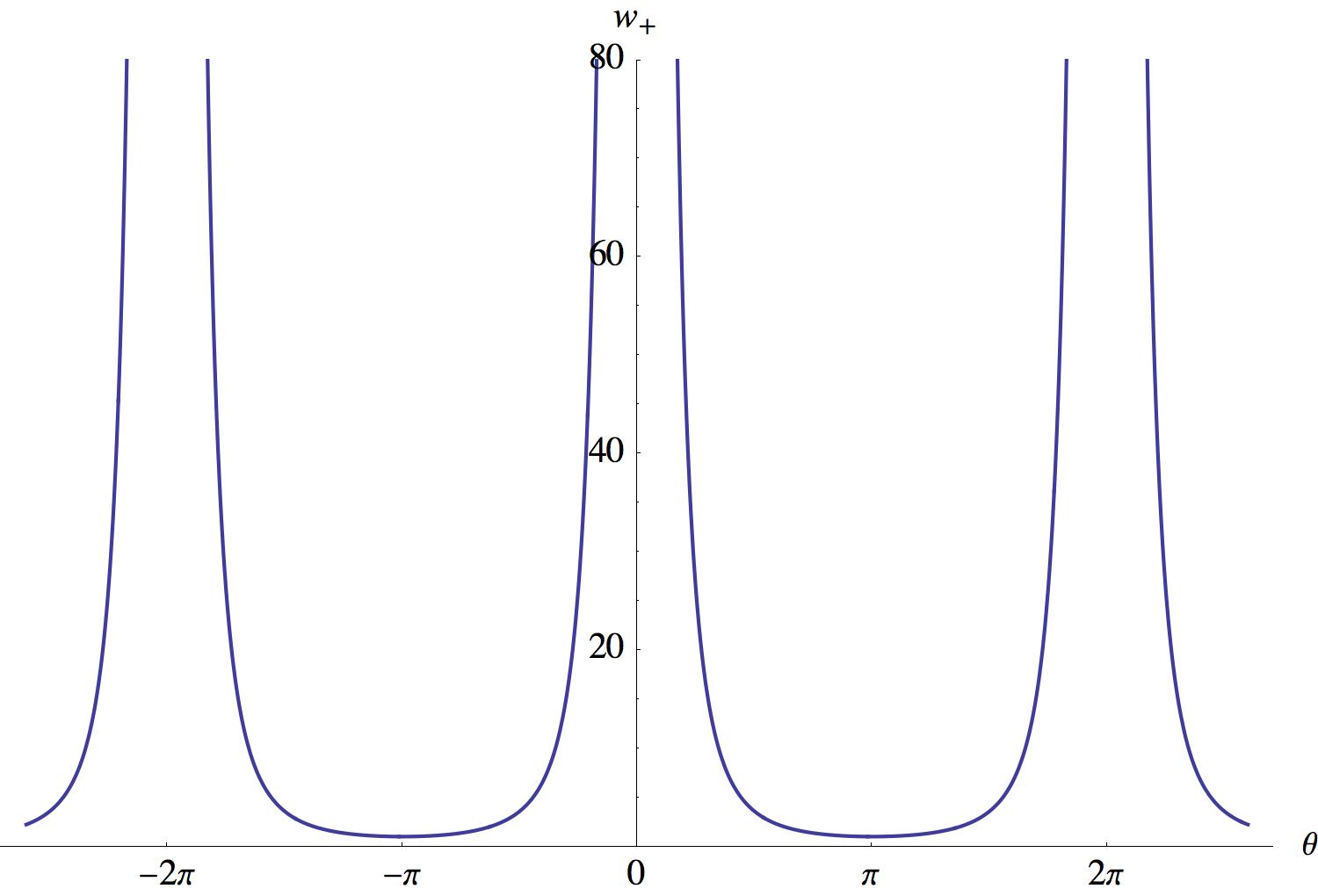}}
\caption{\label{wilson-figure-untwisted}\small The untwisted Wilson loop.}
\end{center}
\end{figure}

The Wilson loops depend on $\theta $ almost trigonometrically, in accord with expectations that $\theta $ is a periodic variable in the dual string picture. However, the dependence on $\theta $ is not entirely analytic, for instance the untwisted Wilson loop diverges as $1/|\theta |^3$ when $\theta$ approaches zero, or any integer multiple of $2\pi $  (fig.~\ref{wilson-figure-untwisted}).  The singularity signals the breakdown of the string description and happens precisely where the gauge theory becomes weakly coupled.

\subsection{Decoupling limit}

We can explore the vicinity of the singular point by considering the limiting case of $\lambda _1\gg\lambda _2$, still assuming $\lambda _2\gg 1$.  This can be called the supergravity decoupling limit to distinguish it from the true decoupling where $\lambda _2\rightarrow 0$. 
All the above formulas then apply with $\theta $ approaching $2\pi $. The effective coupling in this limit coincides with the smaller one:
\begin{equation}
 \lambda \simeq 2\lambda _2,\qquad \theta \simeq 2\pi \left(1-\frac{\lambda _2}{\lambda _1}\right).
\end{equation}
The Wilson loop of the weaker-coupled gauge group stays finite:
\begin{equation}
 W_2\simeq \frac{\,{\rm e}\,^{\sqrt{2\lambda _2}}}{3\cdot 2^{\frac{1}{4}}\pi ^\frac{1}{2}\lambda _2^\frac{3}{4}}\,,
\end{equation}
while the stronger-coupled one diverges as $\lambda _1^3$:
\begin{equation}\label{W1-dec}
 W_1\simeq \frac{\,{\rm e}\,^{\sqrt{2\lambda _2}}}{2^{\frac{1}{4}}\pi ^{\frac{5}{2}}\lambda _2^\frac{15}{4}}\,\lambda _1^3.
\end{equation}

The limiting expression for $W_1$ resembles the SQCD Wilson loop (\ref{WSQCD}) but does not coincide with it in all the detail. The cubic scaling with $\lambda _1$ is reproduced, but the log-suppression is missing and the coefficient of proportionality still depends on $\lambda _2$. The limit $\lambda _{1,2}\rightarrow \infty $, $\lambda _2/\lambda _1\rightarrow 0$, accessible from supergravity, is thus different from the true decoupling where $\lambda _1$ is fixed and  $\lambda _2\rightarrow 0$ (it is enough to take  $\lambda _2\sim 1$). 

It is actually easy to understand why the limits do not commute. The endpoints of the eigenvalue distributions in the supergravity limit behave as
\begin{eqnarray}
 \mu _1&\simeq &\frac{\sqrt{2\lambda _2}}{2\pi }+\frac{2}{\pi }\,\ln\lambda _1
 -\frac{2}{\pi }\,\ln\frac{\pi ^2}{2}
\nonumber \\
\mu _2&\simeq &\frac{\sqrt{2\lambda _2}}{2\pi }-\frac{1}{\lambda }\,\ln 2.
\end{eqnarray}
 Upon true decoupling (in SQCD), one gets \cite{Passerini:2011fe}
\begin{equation}
 \mu _{\rm SQCD}\simeq \frac{2}{\pi }\ln\lambda _1-\frac{1}{\pi }\ln\ln\lambda _1+\,{\rm const}\,,
\end{equation}
again very similar to $\mu _1$, but different in detail. 

The logarithmic growth with $\lambda _1$ in the supergravity limit is an endpoint effect, we still assume that the background, bulk density is a Wigner distribution with a parametrically large width of order $\sqrt{\lambda _2}$, and in particular $\sqrt{\lambda _2}\gg \ln\lambda _1$. Likewise, $W_1$ in (\ref{W1-dec}) depends on $\lambda _1$ through a prefactor, on the background of the leading exponential behavior controlled by $\sqrt{\lambda _2}$. In SQCD, on the contrary, $\ln\lambda _1$ is the largest scale. As $\lambda _2$ decreases, both $W_1$ and $\mu _1$ decrease and should settle to their SQCD values at $\lambda _2\sim 1$. Large logs, $\ln \lambda _1$ and $\ln\ln\lambda _1$, should arise as a remnant of the transitory regime where $\sqrt{\lambda _2}$ and $\ln \lambda _1$ are equally important. 

It is instructive to see what happens to the densities in the decoupling limit.
The gap between the endpoints $\mu  _1$ and $\mu _2$  grows large when $\lambda _1\gg\lambda _2$. Indeed  $\Delta \rightarrow \infty $ as $\theta \rightarrow 2\pi $, which means that $\rho _1$ acquires a long tail extending parametrically far beyond the Wigner distribution. The functional shape of the tail is given  by (\ref{f12-omega}) with $\Delta \rightarrow \infty $:
\begin{equation}
 f_1(\omega )\stackrel{\Delta \rightarrow \infty }{\simeq }
 \frac{i^{\frac{3}{2}}B\left(\frac{1}{2}-\frac{i\omega }{2\pi }\,,\,\frac{1}{2}\right)
 B\left(1-\frac{i\omega }{\pi }\,,\,\frac{1}{2}+\frac{i\omega }{\pi }\right)}{4\sqrt{\pi }\,\omega ^{\frac{3}{2}}}\,\,{\rm e}\,^{i\omega \Delta }.
\end{equation}
The last factor is the Fourier image of a shift operator, as a result $f_1$ becomes effectively a function of $\Delta -\xi $ extending over large distances $\xi \sim \Delta \sim\ln \lambda _1/\lambda _2$.

In the coordinate representation the tail is exponential:
\begin{equation}
 f_1(\xi )\simeq \frac{2\Gamma ^2\left(\frac{3}{4}\right)}{\pi ^{\frac{3}{2}}}\,
 \,{\rm e}\,^{-\frac{\pi }{2}\left(\Delta -\xi \right)},
\end{equation}
or, for the original density,
\begin{equation}
 \rho _1(x)\simeq \frac{2^{\frac{11}{4}}\Gamma ^2\left(\frac{3}{4}\right)}{\pi \lambda _2^\frac{3}{4}}\,\,{\rm e}\,^{\sqrt{\frac{\lambda _2}{8}}-\frac{\pi x}{2}}.
\end{equation}
This is similar but not identical to the asymptotic eigenvalue distribution in SQCD, which at infinite coupling approaches \cite{Passerini:2011fe}:
\begin{equation}
 \rho _{\rm SQCD}(x)\stackrel{\lambda _1=\infty }{=}\frac{1}{2\cosh\frac{\pi x}{2}}
 \simeq \,{\rm e}\,^{-\frac{\pi x}{2}}\,.
\end{equation}
The SQCD eigenvalue density has the same exponential tail but with a different prefactor. Importantly, the behavior at $x\sim 1$ is markedly different: in SQCD the density has a coupling-independent universal shape, while the $\rho _1$  merges with the Wigner distribution at $x\sim \mu _1\sim \sqrt{\lambda _2}$.

\section{Conclusions}

We have studied the expectation value of the circular Wilson loop in the superconformal quiver CFT at strong coupling, starting with the localized partition function on $S^4$. The circular loop is not the only observable accessible via localization. Other marked examples are Wilson loops in higher representations \cite{Fraser:2011qa,Fraser:2015xha}, correlation functions of local operators \cite{Gerchkovitz:2016gxx,Rodriguez-Gomez:2016ijh,Baggio:2016skg,Pini:2017ouj,Beccaria:2018xxl}, correlators between local operators and a Wilson loop \cite{Billo:2018oog,Beccaria:2018owt} and the Bremsstrahlung function \cite{Fiol:2015spa,Mitev:2015oty,Gomez:2018usu}, all potentially calculable by similar methods.

The results for the circular loop are qualitatively consistent with the dual  string picture.
The coupling constant dependence comes out mostly trigonometric, in line with interpretation of $\theta $ as a theta-angle in the string sigma-model, the b-flux through the vanishing cycle of the $AdS_5\times (S^5/\mathbbm{Z}_2)$ orbifold. 
In view of the recent progress on similar problem in $AdS_5\times S^5$ \cite{Forini:2015bgo,Faraggi:2016ekd,Forini:2017whz,Cagnazzo:2017sny,Medina-Rincon:2018wjs},
a more precise, quantitative comparison may actually be within reach. We will not attempt to set up the string calculation here, but will make some general remarks on its salient features. 
 
One can envisage expanding around the minimal surface for the circle, which is an $AdS_2$ hemi-sphere embedded in $AdS_5$ and sitting at a single point on $S^5$ exactly on the orbifold locus. Quantum fluctuations of the string explore the tangent plane to $S^5$ which  in the quiver theory becomes the $\mathbbm{R}\times \mathbbm{C}^2/\mathbbm{Z}_2$ orbifold. The effective string description of the circular Wilson loop is thus a partially massive theory on $AdS_2$ whose massless sector is the $\mathbbm{R}\times \mathbbm{C}^2/\mathbbm{Z}_2$ orbifold. Massive modes originate from fluctuations in $AdS_5$ and presumably cancel once the Wilson loop is normalized to its $\mathcal{N}=4$ value. In all the likelihood the normalized expectation value (\ref{wpm}) is the ratio of the orbifold partition functions on $AdS_2$ at different values of the b-flux:
\begin{equation}
 w_+(\theta )=
 \lim_{\epsilon \rightarrow 0}\frac{Z_{(\mathbbm{C}^2/\mathbbm{Z}_2)_{\epsilon ,\theta }}}{Z_{(\mathbbm{C}^2/\mathbbm{Z}_2)_{\epsilon ,\pi}}}\,,
\end{equation}
where $\epsilon $ is the blowup  parameter that regularized the orbifold geometry. 

The orbifold partition function is naturally represented by an instanton sum:
\begin{equation}
 Z_{(\mathbbm{C}^2/\mathbbm{Z}_2)_{\epsilon ,\theta }}
 =\sum_{k}^{}\mathcal{A}_k\,{\rm e}\,^{-\sqrt{\lambda }\,\epsilon |k|+ik\theta }.
\end{equation}
At finite resolution the instantons are exponentially suppressed but the suppression disappears in the orbifold limit, in accord with our findings. However, an attempt  to extract individual instanton amplitudes from (\ref{wpm}) runs into problems because of the divergences at $\theta = 0$ and $2\pi $. While we understand the origin of these divergences, it is unclear how to regularize them. The principal value prescription does not work, for example\footnote{It does not work for the untwisted Wilson loop. The twisted Wilson loop is analytic in $\theta $ and the principal-value prescription should work.}. The theory at $\theta =2\pi$ has $\lambda _2\sim \mathcal{O}(1)$ and is no longer strongly coupled, even if $\lambda _1\gg 1$. It would be very interesting to make the above arguments more precise and to see how the divergences are resolved (or how they arise) in string theory.
 
\subsection*{Acknowledgements}
We would like to thank C.~Bachas, R.~Klabbers, I.~Klebanov, T.~McLoughlin, C.~Nunez, H.~Ouyang, A.~Parnachev, E.~Pomoni and A.~Tseytlin for interesting discussions and D.~Medina-Rincon for comments on the manuscript. This work was supported by the grant "Exact Results in Gauge and String Theories" from the Knut and Alice Wallenberg foundation and by RFBR grant 18-01-00460 A. 

\appendix

\section{Function $Q$}

The function defined in (\ref{Q-function}) admits an integral representation:
\begin{equation}\label{integral-Q}
 Q(\alpha ,\beta ;q)=\int_{0}^{1}dt\,t^{\beta -1}(1-t)^{\alpha -1}(1+qt)^{-\alpha },
\end{equation}
The only singularities of $Q$ in the finite part of the complex plane are simple poles at non-positive integer $\alpha $. Analyticity in $\alpha $ for $\mathop{\mathrm{Re}}\alpha >0$ easily follows from the integral representation.

It is also easy to develop asymptotic expansions at small and large $\alpha $.
At large positive $\alpha $, 
\begin{equation}\label{large-alpha-Q}
 Q(\alpha ,\beta ;q)\stackrel{\alpha \rightarrow +\infty }{=}
 \frac{\Gamma (\beta )}{\alpha ^\beta (1+q)^\beta }+\mathcal{O}\left(\frac{1}{\alpha ^{\beta +1}}\right).
\end{equation}
At small $\alpha$,
\begin{equation}\label{small-alpha-Q}
 Q(\alpha ,\beta ;q)\stackrel{\alpha \rightarrow 0 }{=}
 \frac{1}{\alpha }-\ln(1+q)-\psi (\beta )-\gamma +\mathcal{O}(\alpha ).
\end{equation}

\bibliographystyle{nb}

\end{document}